\documentclass[prl,preprint,aps,longbibliography,groupedaddress]{revtex4-1}

\usepackage{bbm}
\usepackage{graphicx}
\usepackage{subfigure}
\usepackage{amsmath}
\usepackage{booktabs}

\begin{document}

\title{Indications of superconductivities in blend of variant apatite and covellite}

\author{Hongyang Wang$^{1}$\footnote{\url{wanghy@ipe.ac.cn}}, Yijing Zhao$^{2}$, Hao Wu$^{3}$, Ling Wang$^{4}$, Zhixing Wu$^{5}$, Zhihui Geng$^{6}$, Jiewen Xiao$^{7}$, Weiwei Xue$^{8}$, Shufeng Ye$^{1}$, Ning Chen$^{9}$, Xianfeng Qiao$^{10}$, and Yao Yao$^{10,11}$\footnote{\url{yaoyao2016@scut.edu.cn}}}

\address{$^1$ State Key Laboratory of Mesoscience and Engineering, Institute of Process Engineering, Chinese Academy of Sciences, Beijing 100049, China\\
$^2$ Independent researcher, China\\
$^3$ School of Materials Science and Engineering, Huazhong University of Science and Technology, Wuhan 430074, China\\
$^4$ Beijing Key Laboratory of Ionic Liquids Clean Process, Institute of Process Engineering, Chinese Academy of Sciences, Beijing 100190, China\\
$^5$ Fujian Provincial Key Laboratory of Analysis and Detection Technology for Food safety, College of Chemistry, Fuzhou University, Fuzhou 350108, China\\
$^6$ Graduate School of Engineering, Tokai University, Hiratsuka 2591292, Japan\\
$^7$ College of Environmental Science and Engineering, Beijing Forestry University, Beijing 100083, China\\
$^8$ CAS Key Laboratory of Mechanical Behavior and Design of Materials, Department of Modern Mechanics, University of Science and Technology of China, Hefei 230027, China\\
$^9$ School of Materials Science and Engineering, University of Science and Technology Beijing, Beijing 100083, China\\
$^{10}$ State Key Laboratory of Luminescent Materials and Devices, South China University of Technology, Guangzhou 510640, China\\
$^{11}$ Department of Physics, South China University of Technology, Guangzhou 510640, China}

\date{\today}

\begin{abstract}
Through heavily doping sulfur into an apatite framework, we synthesize a new blend mainly comprising variant apatite and covellite (copper sulfide). Magnetic measurement exhibits that significant diamagnetism appears at around 260~K and drops dramatically below 30~K implying coexistence of two superconducting phases. The upper critical magnetic field is larger than 1000~Oe at 250~K. Electric measurement manifests that the current-voltage curves deviate from the normal linear lineshape suggesting the presence of zero-resistance effect, and the critical current is around 50~$\mu$A at 140~K. These exotic magnetic and electric features strongly indicate these two components, variant apatite and covellite, individually trigger two superconducting phases at near-room and low temperatures.
\end{abstract}

\maketitle

A number of unstable materials in conventional sense serve as research object in modern chemistry. Their thermally stable phase solely occupies a tiny fraction in the whole phase diagram, so precise manipulation of synthesis parameters turns out to be essential \cite{2012prbstable,2012stable,2007stable,2022stable}. The persulfide is a typical instance, which has been theoretically investigated for many years but is difficult to be realized on the experimental side \cite{2012cus,2013cus,2011cusuper,2014cus,2020cus,2020cus2}. The excess sulfur induces a large number of highly active free radicals which benefit the electric conductivity but sensitively influence the thermal stability in ambient circumstance. We thus demand a framework that is able to robustly hold the excess sulfur, and fortunately we have a structure so-called apatite \cite{Lee2,Wang2023,Guo2023,PhysRevB.108.L121110,PhysRevB.108.L201117,PhysRevB.108.235127}. The apatite is a solid framework consisting of a one-dimensional ion channel in which most common anions can be residing. More interestingly, since copper is the most sulfophilic element, the substitution of copper can be largely facilitated by sulfur, otherwise it is difficult to synthesize copper apatite \cite{wang2024unveiling}.

In this context, we uncovered a new concept to synthesize copper persulfide with the help of apatite framework, and the novel magnetic and electric properties of copper-sulfur compounds can thus be comprehensively investigated within this new framework. In our recent works \cite{wang2024observation,wang2024possible,2023lowfield,liu2023longcoherence}, we have concretely discussed the codoping effect of copper and sulfur, manifesting important hint of near-room-temperature superconductivity. However, all the samples reported before were synthesized with sintered apatite being raw materials. The purity can not be ensured during sintering process, so it is quite necessary to utilize low-temperature hydrothermal method in the whole synthesis to mostly keep the chemical valence unchanged. In this paper, we report our new synthesis procedure with the obtained samples exhibiting astonishing superconductivities.

We entirely utilized hydrothermal method under high pressure to synthesize our samples, with the raw materials being copper nitrate, lead nitrate, ammonium phosphate and potassium sulfide. There are two steps. The first step is to synthesize raw lead apatite codoped by copper and sulfur. The mole ratio of Cu$^{2+}$, Pb$^{2+}$, PO$_4^{4-}$, S$^{2-}$ is $9:1:6:2$, and the pH is adjusted to 8 by ammonia water. The mixture is firstly soaked in the water bath at 60~$^{\circ}$C and aged for 24 hours. After sufficiently aging, the color of sample becomes brown and the blue of copper phosphate completely disappears. Afterward the hydrothermal reaction is enabled at 150--160~$^{\circ}$C for 48 hours. After reaction, the sample becomes white gray, and the solution is colorless. Insufficient reaction may lead to wrong colors which means the synthesis is failed. Notice that, if sulfur is absent in this step, the product is lead hydroxyapatite mixed with copper phosphate, and the substitution of copper can not succeed. Then we go to the second step for further sulfur doping. The obtained sample is filtered and dried, and again dispersed in deionized water. The apatite and S$^{2-}$ are mixed again with mole ratio being $1:10$, and the pH is held at 8 by ammonia water. A low pH may lead to the dissociation of apatite. After stirring at 60~$^{\circ}$C for 12 hours, the mixture is hydrothermally treated at 150~$^{\circ}$C for longer than 8 hours. In terms of the color change of the sample, this process may be repeated for two or three times. When sulfur doping is insufficient, the color of solution is yellow meaning the potassium sulfide has not been depleted. A proper doping makes the sample black from the generation of a variant apatite, which is what we want. The overdoping makes it blue and brown, resulting in the complete collapse of apatite framework. After final filtering and drying, the sample has to be purely black without visible metallic luster.

We synthesized two parallel samples labeled by S1 and S2. The sole difference of them is the doping ratio of sulfur in S2 is slightly larger than that in S1. As a comparison, we have also synthesized a lead-free sample labeled by S3, which has been reported in our previous paper \cite{wang2024observation}. Fig.~1 shows the information of structure and element distribution of S1 sample. It is clear from XRD that our sample is a blend of deformed apatite and covellite (copper sulfide). During doping with sulfur, the lattice of raw apatite has been largely shrunk and distorted as the lattice constant is decreased, so we call this new compound as ``variant apatite". The microscopic quantitative results of EPMA in Fig.~1(b) shows the gradual substitution of phosphor and oxygen by sulfur. In particular, in the covellite phase the concentration of sulfur is even around 1.3 times larger than that of copper, indicating there possibly exists copper persulfide. By EPMA WDS mapping as shown in Fig.~1(d)--(h), there are clearly three compounds: Undoped apatite (A), variant apatite (VA), and covellite (C). The effective components turn out to be the latter two.

We employed MPMS-3 SQUID to detect the dc magnetization, as shown in Fig.~2. The magnetization--temperature (MT) curve at 25~Oe with zero-field-cooling (ZFC) measurement exhibits significant diamagnetism below 260~K, and the field-cooling (FC) curve is paramagnetic. The bifurcation between ZFC and FC appears at 250--260~K, which can be regarded as the critical temperature $T_{\rm c}$. The MT curve of S2 sample manifests similar feature in ZFC but the FC magnetization is also diamagnetic. This indicates more sulfur doping in variant apatite can enhance the superconducting features. Interestingly, there is not a plateau-like lineshape at low temperature in both samples, but on the contrary the magnetization quickly decreases following temperature decreasing. It seems we can not observe a pure Meissner phase. In order to comprehend it, the MT curves of S3 sample are recalled as has been reported in our previous paper \cite{wang2024observation}. There are obviously two stages: The first is from 130--230~K that we can regard as the near-room-temperature superconducting phase of variant apatite, and the second is below 30~K which can be recognized as a low-temperature superconducting phase mainly contributed by covellite. For S1 and S2, as the temperature region of superconductivity is larger than that of S3, these two phases mix with each other so that we can not clearly distinguish them. Since the two phases of variant apatite and covellite are associated, we suspect the superconductivity of covellite is induced by that of variant apatite via proximity effect. We also note that the lead-free sample can not be well reproduced with the present synthesis procedure, as the raw material of apatite can not be stabilized without lead.

The magnetization--magnetic field (MH) curves are shown in Fig.~2(d)-(h) at various temperatures. At 250 and 200~K, a linear diamagnetic background is subtracted to display more pretty and featured superconducting hysteresis loops. Essentially, this clear hysteresis has never been observed at this high temperature and ambient conditions in other materials, which can be reasonably recognize as the main feature of near-room-temperature superconductivity. The lower and upper critical magnetic field ($H_{\rm c1}$ and $H_{\rm c2}$) are figured out as well, which basically increases following temperature decreasing. Remarkably, the upper critical field is even close to 1~T at 10~K. Then we measured the MT curves at 1~T and found a notable turning point at around 60~K. Combining our previous reports \cite{wang2024possible}, this might be attributed to the presence of vortex glass.

As shown in Fig.~3(a), the electric transport was measured by four-probe technique on Aglient B2912A and the temperature is controlled by Oxford OptiStatDN. The probes are gilded with the contact diameter being around 0.2~mm, and the distance between them is 2.5~mm. The thickness of samples is 1 or 2~mm. Since the samples are sulfide, both indium and silver electrodes could be sulfurized during measurements. As stated in our previous report \cite{wang2024observation}, this sulfurization remarkably matters in the transport, so we changed the electrodes to be gold in the present work. Since the resistance of samples are extremely small and the critical current is below 100~$\mu$A, it is pretty difficult to measure the transport properties of superconductivity. To this end, the instrument was set to the mode of Compensating Resistance (R Compen). That is, the instrument will automatically initialize an internal voltage bias $U_0$ before the external current is applied, so the readout voltage $U$ is not zero at zero applied current and the readout resistance is calculated by the voltage difference $U-U_0$. It is worth mentioning that, the applied current merely flows through the sample so the voltage drop on the compensating resistance will not be directly influenced by the applied current, meaning that the change of readout voltage can be solely realized as dropping on the sample. We have also used a standard copper sheet with thickness of 40~$\mu$m to calibrate the compensating resistance, as shown in inset of Fig.~3(i). That is, we changed the distance of electrodes and linearly fitted the results to obtain the value at 0~cm. The instrument will also automatically shift the measure range at 106~$\mu$A and below 20~$\mu$A the noise is too strong, so we always kept our measurement within 20--100~$\mu$A to ensure sufficient precision. Furthermore, as there are a number of free ions and defects in the bulk, before each measurement an electric biasing with large current was enabled to fill the defects. All the IV curves are averaged over 100 samplings and RT curves are over 2000 runs.

Although there is a very large compensating resistance from the instrument over which the initial voltage drops, the appearance of voltage plateau in current--voltage (IV) curves, i.e. following applied current increasing the voltage on the sample does not change, can be safely regarded as from quantum transport, namely zero-resistance effect. Exceeding the so-called critical current $I_{\rm c}$, the voltage is onset and quickly rises and then turns to the normal linear IV curve in the normal metal state, implying the transition from superconductivity to normal phase. At low temperatures, the normal state appears above 100~$\mu$A which can not be detected in our present experiment. We can see from Fig.~3(b)--(g) that, from 140--240~K this typical superconducting IV curves are clearly observed, and the critical current basically decreases with temperature increasing and can be roughly fitted by a quadratic function. At 260~K, the voltage plateau is completely invisible suggesting the critical current becomes below 20~$\mu$A, or even disappears. Interestingly, the IV curve at 80~K exhibits multiple strange plateau-like voltage regions which are also observed at other temperatures below 130~K (not shown). Combining the magnetic properties reported above, we realize this can be related to the coexistence of superconductivities in both variant apatite and covellite. Superconducting proximity effect of the variant apatite induces the superconductivity of covellite, so that the conductivity is largely enhanced in the main phase of covellite.

We have also measured the resistance--temperature (RT) curves of S1 and S2 samples. Subtracting the compensating resistance, the resistance of sample is 1--3~m$\Omega$ at room temperature, meaning that the resistivity is in the order of $10^{-6}$--$10^{-7}~\Omega\cdot m$. Considering the short electrode distance in our setup and the large fluctuation of initial voltage $U_0$, the resistance is too small to be precisely measured, so it is difficult to determine critical temperature by RT curves. By exerting large current using PPMS, as reported by our previous paper \cite{wang2024observation}, the RT curve exhibits a turning point at around 230~K, which can be realized as a hint of second-order phase transition. Based on the present experiment, it is then reasonable to estimate from our results that, the resistivity below 240~K is at least in the order of $10^{-8}~\Omega\cdot m$ or even smaller. This indicates the resistivity of our sample is at least comparable to or even smaller than that of copper, strongly suggesting we have got zero-resistance effect.

Before ending, we would like to discuss more on the reproduction and cherry-picking of samples. The whole reaction process of the synthesis is a nonequilibrium dynamic process. Following continuous doping of sulfur, the apatite in the raw material will gradually evolve to covellite. We can divide the process into five steps and in each step there is a main effective component in the blend: Raw material of apatite, sulfur doped apatite, variant apatite as main phase blended with covellite, covellite as main phase doped with variant apatite, and covellite. S1 and S2 samples belong to the third and fourth category, respectively. We have also synthesized three other samples, S4, S5 and S6, individually terminated at relevant steps to show the change during the variant process. The XRD of them can be seen in Fig.~4(a), and from Fig.~4(b) it is clear that during the reaction the lattice is continuously shrunk and distorted. The doped apatite basically holds the main framework of apatite, though the lattice distortion has been very strong. In the overdoped phases including variant apatite and covellite the framework gets close to collapse. The EPMA of covellite exhibits the main phase is indeed copper sulfide and also some persulfide particles are observed. Although both variant apatite and copper persulfide are unstable by themselves in ambient conditions, the distortion of lattice stabilizes them implying the association between them. The magnetization of additional samples are shown in Fig.~4(c)--(e). The undoped apatite manifests paramagnetism, and the covellite has no magnetic response except below 30~K. The doped apatite and covellite below 30~K exhibit diamagnetism which is what we want. Consequently, the most important component, variant apatite, is generated as an intermediate in step 2, 3 and 4. In order to optimize the synthesis, therefore, the manipulation of doping and defect concentration plays the essential role.

In summary, based on our novel synthesized blend of variant apatite and covellite, we observe near-room-temperature superconductivity coexisting with another low-temperature superconducting phase. Following our concept of synthesis, we believe the properties of superconductivities can be largely improved in the near future and promising applications can be soon established.

\section{Acknowledgments}

The authors gratefully acknowledge support from the National Natural Science Foundation of China (Grant Nos.~12374107 and 52304430). H. W. thanks Dr. Jicheng Liu for fruitful discussions on the synthesis and structure analysis.

\section{Author contributions}

Hongyang Wang: Methodology, Conception, Supervision, Synthesis, Characterization, Magnetic measurement, and Writing - Original Draft.

Yijing Zhao: Formula design, and Synthesis for sample S2.

Hao Wu: Electric measurement, and Magnetic measurement.

Ling Wang: Magnetic measurement, and Device support.

Zhixing Wu: Resource and Data Curation.

Zhihui Geng: Investigation.

Jiewen Xiao: Device support.

Weiwei Xue: Device support.

Shufeng Ye: Funding.

Ning Chen: Formal analysis, and Writing - Review \& Editing.

Xianfeng Qiao: Electric measurement, Software, and Data Curation.

Yao Yao: Formal analysis, Validation, Supervision, Electric measurement, Data Curation, Visualization, and Writing - Original Draft, Review \& Editing.

\bibliography{Covellite_final.bbl}

\begin{figure*}[t]
    \centering
    \includegraphics[width=1.0\linewidth]{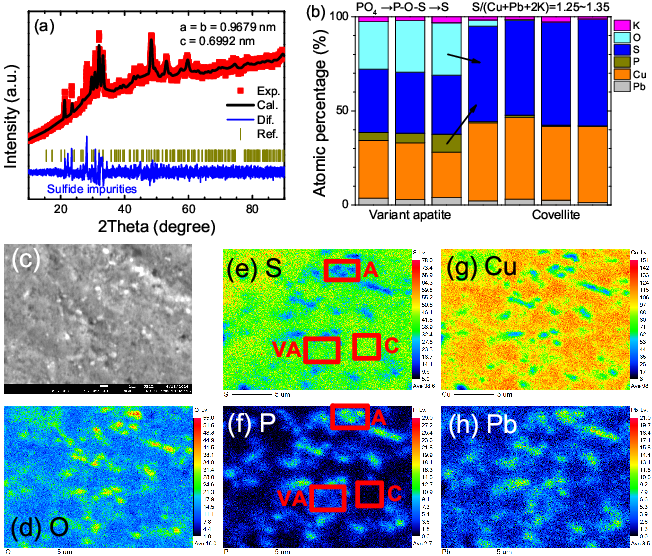}
    \caption{(a) XRD spectrum of S1 sample. The difference indicates the presence of sulfide impurities. The lattice constants are shorter than that of apatite implying it is distorted. (b) EPMA quantified atomic percentage on seven spots. Arrows indicate phosphor and oxygen are substituted by sulfur, suggesting the reaction process from apatite to covellite. (c) SEM image of S1 sample. The white spots are variant apatite and the gray regions are covellite. (d)--(h) EPMA mapping of individual elements: oxygen, sulfur, phosphor, copper and lead. The red squares figure out three main phases in the sample: Undoped apatite (A), variant apatite doped by sulfur (VA) and copper persulfide (C). }
\end{figure*}

\begin{figure*}[t]
    \centering
    \includegraphics[width=0.7\linewidth]{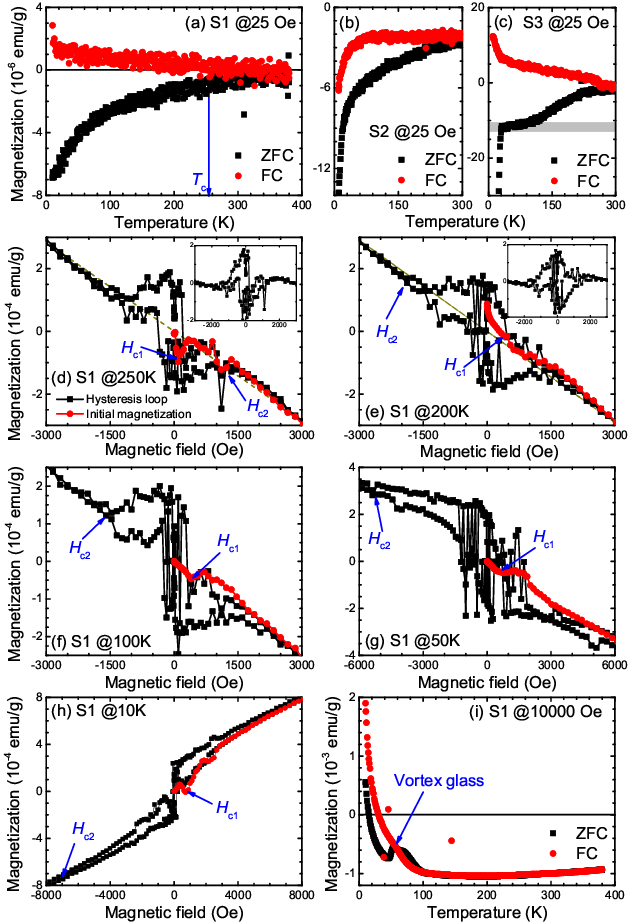}
    \caption{(a)--(c) MT curves of ZFC and FC measurements of S1, S2 and S3 samples at 25~Oe. The critical temperature is within 250--260~K. The plateau-like region of magnetization of S3 is highlighted by gray shade. (d)--(h) MH hysteresis and initial magnetization curves at five temperatures. The lower and upper critical magnetic fields $H_{\rm c1}$ and $H_{\rm c2}$ are pointed out respectively. At 250 and 200~K, the diamagnetic background is indicated by yellow dashed lines, and the MH curves subtracting this background are displayed in the insets. (i) MT curves of ZFC and FC of S1 at 10000~Oe, which manifests vortex glass feature at low temperature.}
\end{figure*}

\begin{figure*}[t]
    \centering
    \includegraphics[width=0.6\linewidth]{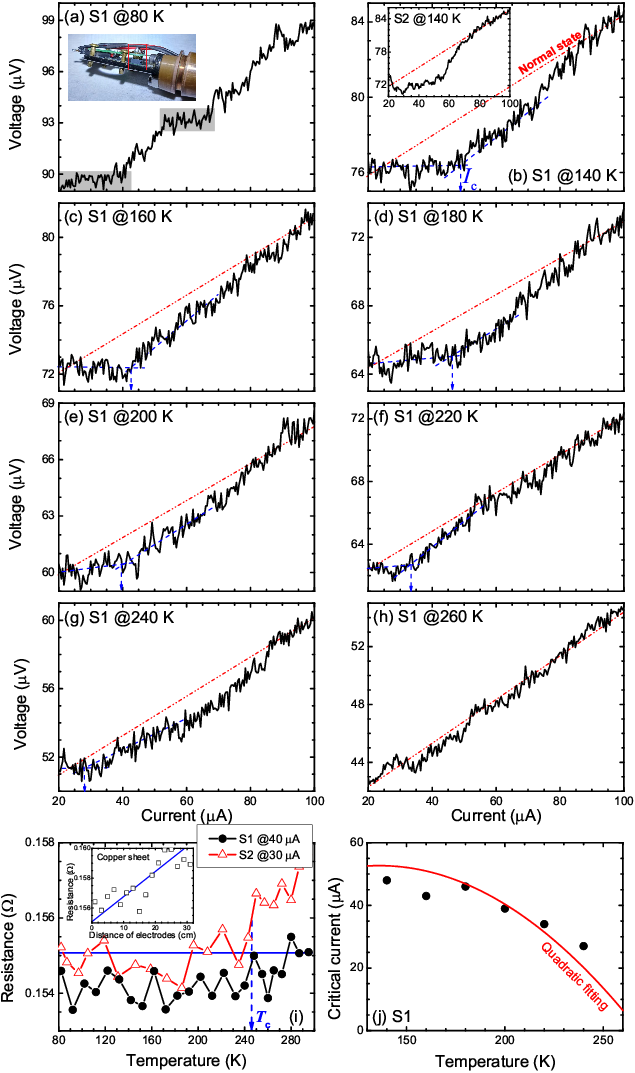}
    \caption{(a)--(h) IV curves of S1 sample at eight temperatures. The two plateau-like voltage regions at 80~K are sketched by gray shades. The red dash-dot lines at other temperatures figure out linear IV curves in normal metal state. Blue dashed lines and arrows indicate plateau-like and onset region and the relevant critical currents $I_{\rm c}$. Inset in (a) shows the photo of our four-probe setup. Inset in (b) displays the relevant result of S2 sample. (i) RT curves of S1 and S2 at 40 and 30~$\mu$A, respectively. The compensating resistance of the instrument is indicated by blue line, which is calibrated by copper sheet with various electrodes distances as displayed in the inset. (j) The estimated critical current versus temperature which is quadratically fitted.}
\end{figure*}

\begin{figure*}[t]
    \centering
    \includegraphics[width=1.0\linewidth]{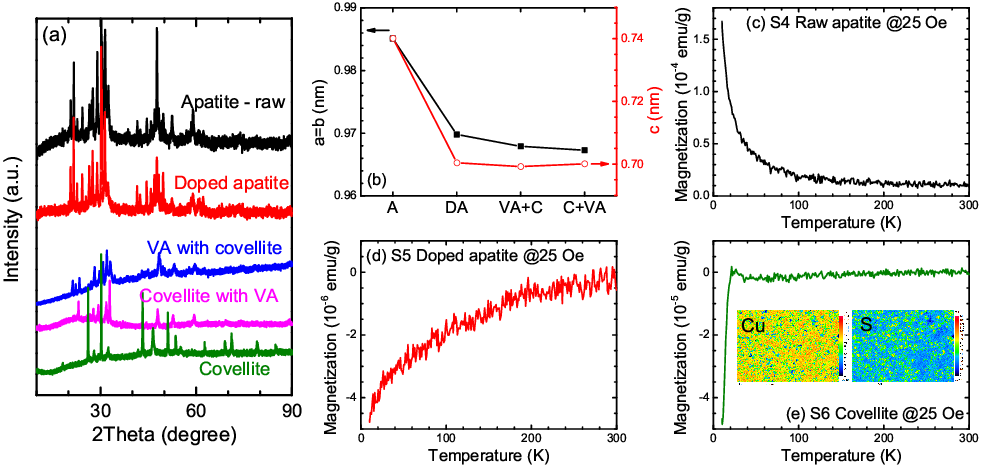}
    \caption{(a) XRD of five samples obtained in each step during the whole reaction process: Apatite - raw, sulfur doped apatite, variant apatite as main phase blended with covellite, covellite as main phase with variant apatite, and covellite. (b) The lattice constants show the shrink of the whole structure in the reaction. (c)--(e) MT curves at 25~Oe of three additional samples S4, S5 and S6, which are raw apatite, doped apatite and covellite, respectively. Inset in (e) displays the distribution of copper and sulfur in S6, indicating the main phase is copper sulfide and there are also many spots of copper persulfide. }
\end{figure*}

\end{document}